\documentclass[sn-mathphys,Numbered]{sn-jnl}

\usepackage{graphicx}%
\usepackage{multirow}%
\usepackage{amsmath,amssymb,amsfonts}%
\usepackage{amsthm}%
\usepackage{mathrsfs}%
\usepackage[title]{appendix}%
\usepackage{xcolor}%
\usepackage{textcomp}%
\usepackage{manyfoot}%
\usepackage{booktabs}%
\usepackage{algorithm}%
\usepackage{algorithmicx}%
\usepackage{algpseudocode}%
\usepackage{array}
\usepackage[caption=false,font=normalsize,labelfont=sf,textfont=sf]{subfig}
\usepackage{textcomp}
\usepackage{stfloats}
\usepackage{url}
\usepackage{verbatim}
\usepackage{listings}%
\usepackage{graphicx}
\usepackage{tabularx} 
\usepackage{multirow} 
\usepackage{hhline} 
\usepackage{lipsum}
\usepackage[normalem]{ulem}
\usepackage{soul}

\theoremstyle{thmstyleone}%
\theoremstyle{thmstyletwo}%

\theoremstyle{thmstylethree}%

\begin{document}


\title[Article Title]{A low latency attention module for streaming self-supervised speech representation learning}


\author*[]{\fnm{Jianbo} \sur{Ma}}\email{jianbo.ma@dolby.com}
\author[]{\fnm{Siqi} \sur{Pan}}
\author[]{\fnm{Deepak} \sur{Chandran}}
\author[]{\fnm{Andrea} \sur{Fanelli}}
\author[]{\fnm{Richard} \sur{Cartwright}}

%

\affil[]{\orgname{Dolby Laboratories}}


\abstract{The transformer is a fundamental building block in deep learning, and the attention mechanism is the transformer's core component. Self-supervised speech representation learning (SSRL) represents a popular use-case for the transformer architecture. Due to transformers' acausal behavior, the use of transformers for SSRL has been predominantly focused on acausal applications. However, several media processing problems, such as speech processing, require real-time solutions. In this paper, we present an implementation of the attention module that enables training of SSRL architectures with low compute and memory requirements, while allowing real-time inference with low and fixed latency. The attention module proposed in this paper includes two components, streaming attention (SA) and low-latency streaming attention (LLSA). The SA represents our proposal for an efficient streaming SSRL implementation, while the LLSA solves the latency build-up problem of other streaming attention architectures, such as the masked acausal attention (MAA), guaranteeing a latency equal to one layer even when multiple layers are stacked. We present a comparative analysis between the vanilla attention, which we will refer here as acausal attention (AA), the SA, and the LLSA, by training a streaming SSRL with automatic speech recognition as downstream task. When training on librispeech-clean-100 and testing on librispeech-test-clean, our low-latency attention module has a word error rate (WER) of 5.84\%, which represents a significant improvement over the MAA (WER = 13.82\%). Our implementation also reduces the inference latency from 1.92 to 0.16 seconds. The proposed low-latency module preserves many of the benefits of conventional acausal transformers, but also enables latency characteristics that make it applicable to real-time streaming applications. 
}


\maketitle

\section{Introduction}\label{sec:introduction}

{T}{he} transformer, introduced in \cite{vaswani2017attention}, is one of the most popular building blocks in modern neural network architectures. Transformers have been applied to many fields, such as natural language processing (NLP)~\cite{vaswani2017attention}\cite{wolf2020transformers}, natural language understanding (NLU) \cite{jiao2019tinybert}, computer vision (CV) \cite{khan2022transformers}, and speech and audio processing~\cite{wang2020transformer}\cite{moritz2020streaming}. The vanilla transformer is characterized by one or more attention units, which are inherently acausal, since they use all of the information available in a sequence of data to produce a new output. In audio applications, the acausality of the attention model prevents the use of the vanilla transformer in real-time or streaming applications. Within the context of this paper, we adopt the term acausal attention (AA) to denote the vanilla attention mechanism. Furthermore, the term 'acausal' is employed to characterize the requirement of accessing future information for processing data at the current position, such as the time or index of sequential data. In contrast, the term 'causal' is used to signify the absence of this requirement, indicating a processing approach that does not rely on future information.

Several authors have proposed methods for creating causal transformers. For example, the chunk-wise transformer \cite{zhang2020streaming}\cite{chiu2017monotonic}\cite{li2020comparison} segments the input vectors into sequential chunks and applies the attention mechanism within each chunk. While this process overcomes the acausality of transformers, it treats each chunk independently, causing sample discontinuity at the edges of each chunk. The memory-based method introduced by Wu et. al. \cite{wu2020streaming}\cite{shi2021emformer} uses an additional memory bank to enable longer contextual dependency, but careful segmentation is required. Another popular group of methods involves masking or time-restricting the attention score to limit the receptive field \cite{vaswani2017attention}\cite{chen2021developing}\cite{povey2018time}\cite{moritz2020streaming}\cite{tripathi2020transformer}\cite{huang2023self}. However, while those approaches solve the causality problem and guarantee fixed latency attention, they all rely on some sort of masking, which is a computationally inefficient technique, particularly on audio data. Moreover, as discussed in Section \ref{sec:lowLatencyStreamingAttentionKernel}, these methods also suffer from latency build-up as multiple restricted attention layers with look-ahead are concatenated.

Self-supervised speech representation learning (SSRL) has proven to be a successful strategy to learn generic representations of speech, with transformer architectures being widely adopted for its implementation. Transformers have mainly been applied to acausal SSRL applications, due to their aforementioned limitations.  While this is not a problem when training and testing on short audio sequences (e.g., recordings of single speech utterances), it limits the transformer's usage in streaming applications where causality and fixed latency is required, such as telecommunication and broadcast. For this reason, the implementation of a low-latency attention model is of particular interest. On top of that, a memory and computationally efficient attention module is highly desired, because SSRL typically requires large amount of training data and is computationally demanding at training time. 

SSRL learns generic representations from unlabeled audio data and can be used for a range of downstream tasks \cite{mohamed2022self}. The phase of learning generic representations is called pre-training and the phase that adapts to a downstream task is referred as fine-tuning. Different types of SSRL have been proposed. For example, \cite{chung2019unsupervised} proposed the Autoregressive Predictive Coding (APC) which uses previous frames to predict the next frames. While APC only uses the previous frames, DeCoAR \cite{ling2020deep} and ELMo  \cite{peters2018deep}\cite{mohamed2022self} use frames from both directions. In addition to reconstruct the speech features, other methods like wav2vec~\cite{schneider2019wav2vec} and wav2vec 2.0~\cite{baevski2020wav2vec} use the Contrastive Predictive Coding (CPC)~\cite{oord2018representation}, where both positive and negative samples are drawn from the latent space conditioned on contextual vector to calculate the InfoNCE~\cite{oord2018representation}. More recent methods use the masked prediction techniques~\cite{huang2023self}\cite{mohamed2022self}\cite{hsu2021hubert}, inspired from the masked language model in BERT~\cite{devlin2018bert}\cite{chiu2022self}\cite{chen2022beats}. The development of those techniques are mainly focused on the design of proxy tasks that enforce self-supervision. Those SSRL models typically use the acausal transformer as the backbone, limiting their application to offline use-cases. The lack of an attention mechanism that is computational efficient while retaining low latency have limited the application of SSRL to real-time applications.  

In this paper, we propose a low-latency attention module to overcome the limitations described above. Our low-latency attention module contains two components. The first component, which we will refer to as the {\it Streaming Attention} (SA), restricts the receptive field to guarantee higher computational and memory efficiency than previous methods. The second component, which we will refer to as the {\it Low Latency Streaming Attention} (LLSA), builds on SA and solves the latency build-up problem as layers are concatenated, facilitating low-latency use in real-time systems. We derive forward- and back-propagation equations for both methods and conduct experiments with dedicated GPU implementations of SA and LLSA. The proposed modules are then applied to a streaming SSRL. Specifically, we use HuBERT~\cite{hsu2021hubert} as the test-bed for a streaming SSRL task since it is one of best performing SSRL models. We name our resulting real-time SSRL model SHuBERT. Compared to previous causal transformer designs, we show theoretically that the SA achieves higher computational and memory efficiency (Section \ref{sec:computational_efficiency}). We also empirically prove that SA achieves high computational efficiency for multiple heads and receptive fields within the attention module \cite{vaswani2017attention}. Working with SHuBERT and an Automatic Speech Recognition (ASR) downstream task, we confirm experimentally that LLSA reduces the latency (from more than 3 seconds to 300 milliseconds) with only a minor drop of performance. The authors believe that the proposed low-latency attention modules SA and LLSA are applicable to a wide range of tasks and applications.

In summary, the contribution of this paper is as follows.
We propose a low latency attention module that achieves computational efficiency and retains low latency. The proposed SA requires less computational memory and is more efficient than the popular masked acausal attention method. The LLSA further solves the latency build-up issue without extra manipulation of input data as~\cite{shi2021emformer}\cite{chen2021developing}. The experiments show streaming SSRL (SHuBERT) with streaming ASR as downstream task achieves state-of-the-art performance with the proposed low latency attention module. We believe the proposed low latency attention module paves the way for future research in order to extend the SSRL to support more real-time scenarios.

\section{Related work}

\subsection{Acausal Attention}
\label{sec:aa}
The structure of a transformer is described in \cite{vaswani2017attention}, where the concept of Multi-Head Attention (MHA) is introduced. We here describe the core building block of the MHA, the Scaled Dot-Product Attention (SDPA) unit, using a vectorial representation (in place of the matrix representations used in \cite{vaswani2017attention}), in order to highlight the temporal relationships.

\begin{figure}
	\centering
	\includegraphics[scale=0.3]{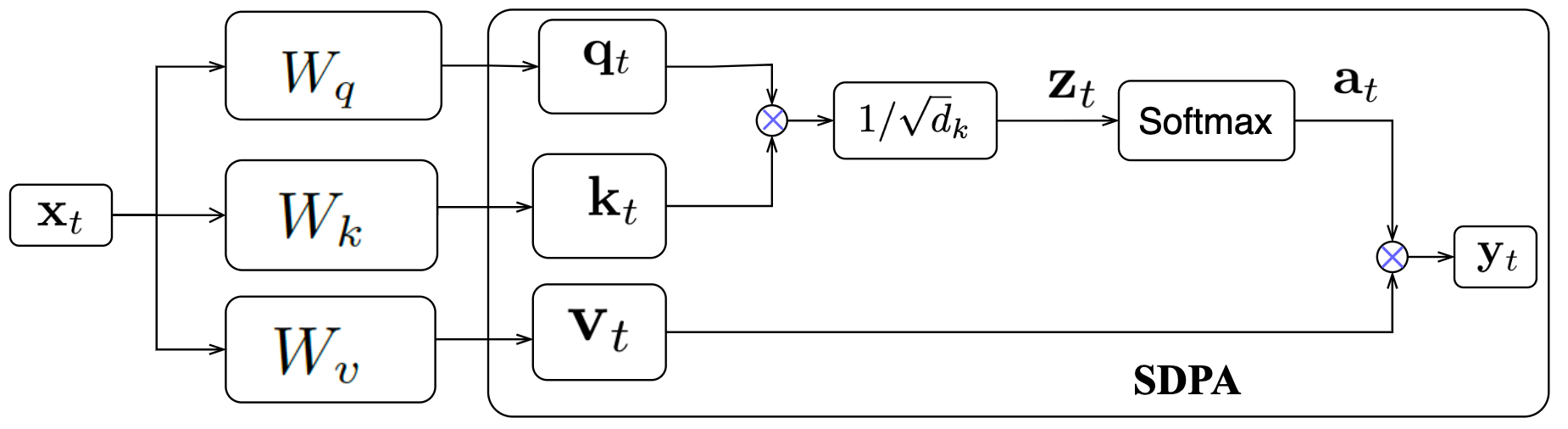}
	\vspace{-5pt}
	\caption{Scaled Dot-Product Attention unit.}
	\label{fig:sdpa}
	\vspace{-10pt}
\end{figure}

Fig.~\ref{fig:sdpa} shows the SDPA unit. To implement self-attention in a transformer, each input vector $\mathbf{x}_{t}$ is projected into three quantities known as the \emph{query} ($\mathbf{q}_{t}$, of length ${d}_{k}$), the \emph{key} ($\mathbf{k}_{t}$, also of length ${d}_{k}$) and the \emph{value} ($\mathbf{v}_{t}$, of length ${d}_{v}$) by multiplying it with $\mathrm{W}_{q}$, $\mathrm{W}_{k}$, $\mathrm{W}_{v}$ respectively. These projections can be implemented causally since they rely only on input from time $t$. $\mathbf{q}_{t}$, $\mathbf{k}_{t}$ and $\mathbf{v}_{t}$ are then formed by equally splitting the output of projections into the number of heads. Together, $\mathbf{q}_{t}$, $\mathbf{k}_{t}$ and $\mathbf{v}_{t}$ form the input to the SDPA unit. Note that Fig.~\ref{fig:sdpa} only shows one head of MHA.

In the SDPA unit, $\mathbf{z}_{t}$, of length ${N}_{T}$, is obtained as the scaled dot-product between the query $\mathbf{q}$ at time index $t$ and each of the keys $\mathbf{k}$, 
\vspace{-4pt}
\begin{equation}
	\label{eq:na-dot-product}
	\mathbf{z}_{t} = [\mathbf{k}_{0},\mathbf{k}_{1},....,\mathbf{k}_{{N}_{T}-1}]^{\intercal} \frac{\mathbf{q}_{t}}{\sqrt{d_{k}}}
	\vspace{-0.15cm}
\end{equation}
where $t$ is the time index, $(\cdot)^\intercal$ is transpose operator and $N_{T}$ is the number of frames. The attention score vector $\mathbf{a}_{t}$ is then computed as  $\mathbf{a}_{t} = softmax(\mathbf{z}_{t})$,
where $softmax(\cdot)$ is the softmax operator. Finally, $\mathbf{y_{t}}$ is computed by multiplying the attention score vector $\mathbf{a_{t}}$ by the values $\mathbf{v}$:
\vspace{-4pt}
\begin{equation}
	\label{eq:na-for-v}
	\mathbf{y}_{t} = [\mathbf{v}_{0}, \mathbf{v}_{1},...,\mathbf{v}_{N_{T}-1}]^{\intercal} \mathbf{a}_{t}.
	\vspace{-0.45cm}
\end{equation}

\subsection{Masked Acausal Attention}
\label{sec:subsec_mask}
As observed in Eq.~\eqref{eq:na-dot-product} and Eq.~\eqref{eq:na-for-v}, all the key and value vectors from time $0$ to $N_{T}-1$ are used when calculating each output $\mathbf{y}_{t}$. Therefore, the SDPA unit is acausal. In \cite{vaswani2017attention} and \cite{chen2021developing}, the authors introduce the idea of masking the attention score $\mathbf{a_{t}}$ to remove dependency on the future of the input. The masking vector is
\vspace{-4pt}
\begin{equation}
	\label{eq:na-mask}
	\mathbf{m}_{t} = [0,0, \dots , 1,1,\dots,1,\dots,0],
	\vspace{-0.15cm}
\end{equation}
where $1$ preserves the value of the corresponding time index when computing $\mathbf{a}_{t}$, and $0$ indicates that the corresponding position in $\mathbf{z}_{t}$ is replaced by a large negative value, so that those values are not used to compute $\mathbf{y}_{t}$. This strategy can also be used to limit the amount of past data that is used in the computation. 

While the output $\mathbf{y}_{t}$ does not depend mathematically on unwanted future or past input, all the $\mathbf{z}_{t}$ corresponding to unwanted input positions are still computed and then subsequently replaced, meaning that the system remains acausal. This method is computationally inefficient especially when applied to audio data with large receptive fields, since all of the $\mathbf{z}_{t}$ values are computed in $O({d}_{k}{N}_{T}^2)$ time, and additional computations are required to replace most of them with a large negative value. Moreover, large amount of memory (from the already scarce GPU availability) must be allocated to store all of the $\mathbf{z}_{t}$ values, many of which will be unused. For those reasons, using the masking mechanism ends up limiting the effective batch sizes that can be used during training and increases the computation time per batch.

Consider the example of a $60$-second speech utterance with feature extraction running on a $10$-ms time step. This vector would have ${N_{T}} = 6000$. If we were to restrict the receptive field of the network to 1.2 seconds, each masking vector $\mathbf{m}_{t}$ would consist of 120 ones and 5880 zeros. This results in the use of only 720,000 out of the 36 million attention values that are computed and stored in memory, as well as in 35,280,000 dummy value replacements.

\subsection{HuBERT}
\label{sec:related_works_hubert}
\begin{figure}[!htb]
	\begin{minipage}{0.43\textwidth}
		\centering
		\includegraphics[width=\linewidth]{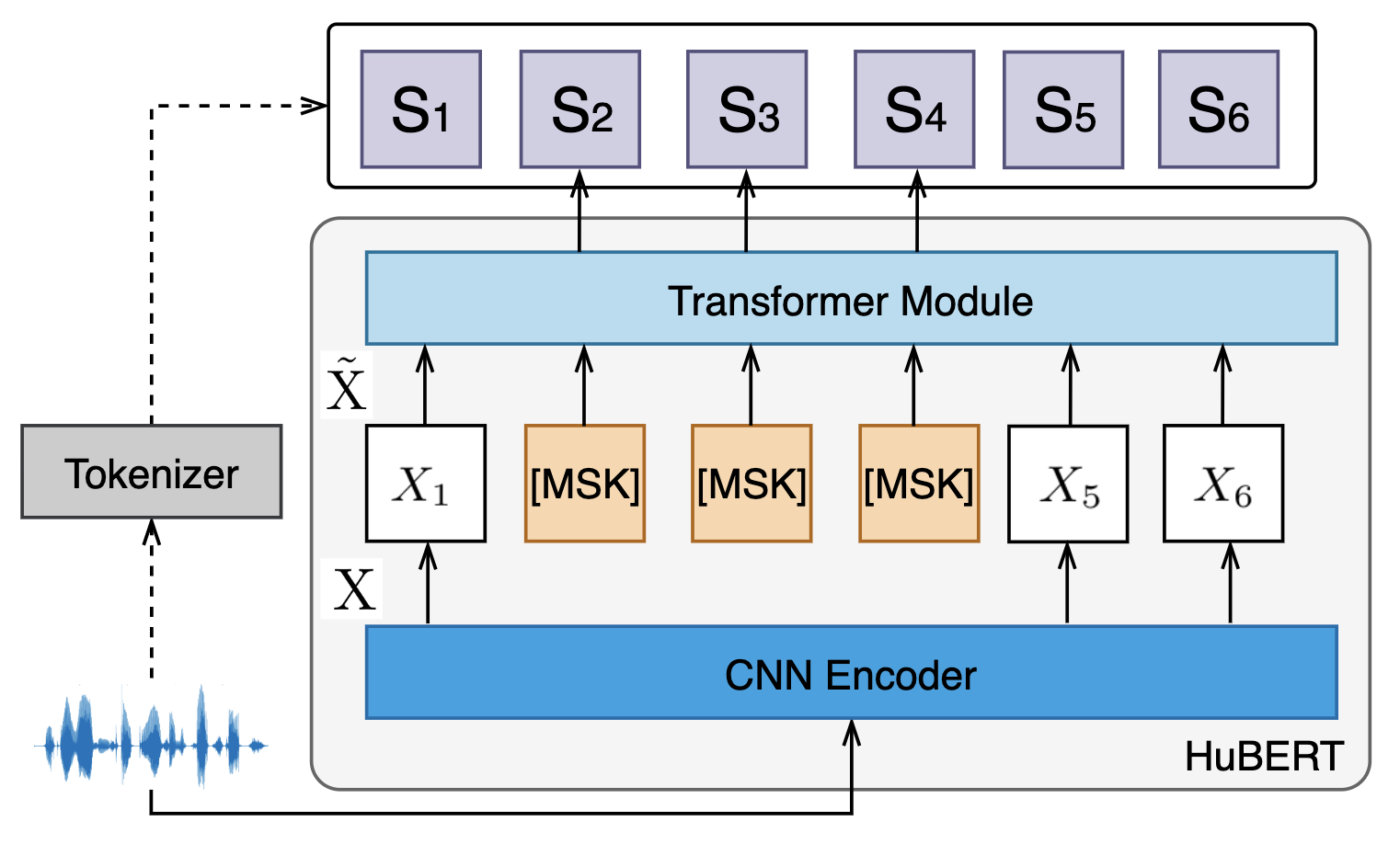}
		\caption{HuBERT Model Architecture~\cite{hsu2021hubert}}\label{fig:hubert}
	\end{minipage}\hfill
	\begin{minipage}{0.43\textwidth}
		\centering
		\includegraphics[width=\linewidth]{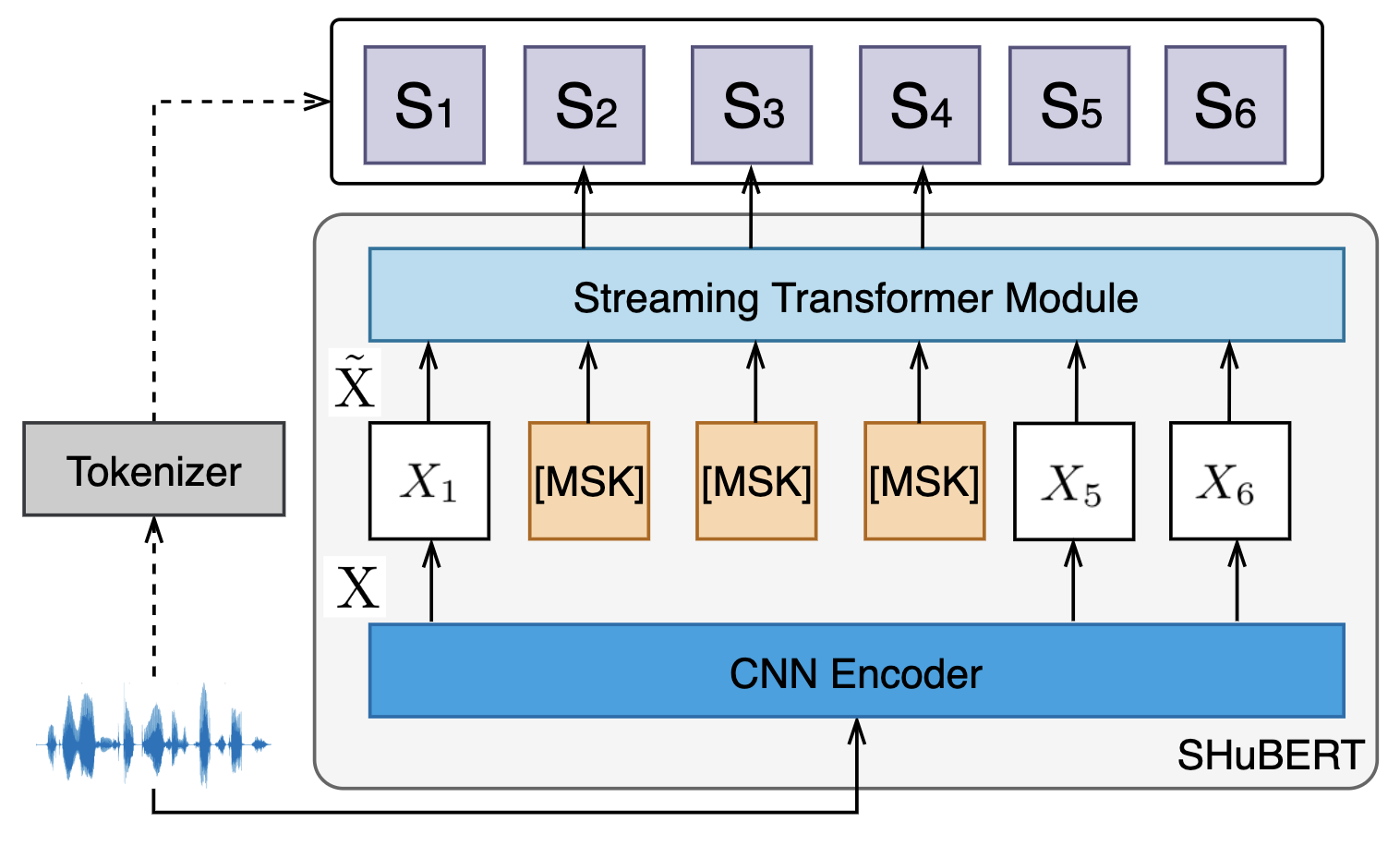}
		\caption{SHuBERT Model Architecture}\label{fig:shubert}
	\end{minipage}
\end{figure}

\noindent The HuBERT architecture \cite{hsu2021hubert} is used as the test bed in this paper. Fig.~\ref{fig:hubert} shows the HuBERT implementation specific to the masked speech prediction approach \cite{huang2023self}\cite{mohamed2022self}, where a tokenizer is used to generate the pseudo labels for each time step $\mathrm{s}_{t}$ of the audio segment. In HuBERT, the tokenizer uses k-means to cluster the MFCC acoustic features during the first training iteration. During the following training steps, features are extracted from hidden layers in the transformer module.  In Fig. \ref{fig:hubert}, $\mathrm{X}$ denotes the acoustic feature generated after the convolutional neural networks (CNN) encoder. Masked embeddings are used to replace the features in masked regions and form the corrupted features $\widetilde{\mathrm{X}}$.  $\widetilde{\mathrm{X}}$ is fed into the transformer block to generate posterior probability $p(s_{t} | \widetilde{\mathrm{X}}, t)$ for each frame. Cross-entropy loss is used and losses are split into masked loss and unmasked loss.

To generate the posterior probability, all embeddings of each time step in $\widetilde{\mathrm{X}}$ are used. This means that the future information is used to generate the current posterior probability, which suggests that the process is acausal. Fig. \ref{fig:latency_illustration}(a) also shows this acausal behaviour: all of the past and future frames (in green) are used to process the current frame (in red). The acausality is due to the SDPA block, as described in Section.~\ref{sec:aa}.

To conduct experiments on the HuBERT model, we built the offline HuBERT following the same training steps described in~\cite{hsu2021hubert}. As shown in Table~\ref{tab:hubert_asr_baselines}, the results are close to the ones reported in~\cite{hsu2021hubert}, with librispeech test-clean and test-other being the test sets. Details are reported in Section~\ref{sec:experimental_results}.

\section{Low latency attention module}
\label{sec:llam}
\noindent The proposed low latency attention module includes streaming attention (SA) and low latency streaming attention (LLSA). The SA improves memory usage, thus enabling efficient training. The LLSA solves the latency build-up problem, thus reducing latency during offline training. In the Section \ref{sec:experimental_results}, we show the efficacy of the proposed method.

\vspace{-0.25cm}
\subsection{Streaming Attention}
\vspace{-0.15cm}
\label{sec:streamingAttentionKernel}
We now introduce Streaming Attention (SA), a method to limit the computation of only the elements of $\mathbf{z}_{t}$ that are required for the desired receptive field. Compared to the masked acausal attention method described above, it achieves substantially higher computational efficiency and much lower memory usage for a given batch size.

The back-propagation algorithm \cite{rumelhart1986learning} is commonly used for training neural networks, and consists in propagating the error through the network \cite{Goodfellow-et-al-2016} and adjusting the model weights toward the minimum error configuration. The chain rule of calculus \cite{Goodfellow-et-al-2016} is used to increase the efficiency of the back-propagation. The derivatives of the loss with respect to each of the SDPA inputs need to be computed to properly update the parameters of the model. This includes $\frac{\partial \ell}{\partial \mathbf{v}_{t}}$, $\frac{\partial \ell}{\partial \mathbf{k}_{t}}$ and $\frac{\partial \ell}{\partial \mathbf{q}_{t}}$, where $\ell$ is the overall loss for one step, corresponding to a mini-batch.
\vspace{-0.3cm}
\subsubsection{Forward Propagation}
\vspace{-0.1cm}
The core idea of SA is to limit the receptive field to $A$ frames in the ``future" (\emph{look-ahead}), and $B$ frames in the past (\emph{look-back}), in relation to the input data at time $t$.  Then, a causal SPDA operator is achieved by introducing $A$ frames latency and a fixed receptive field of $B+1+A$ frames. We introduce $A$ and $B$ into Eq.~\eqref{eq:na-dot-product} to obtain
\begin{equation}
	\label{eq:dot-product}
	\mathbf{z}_{t} = [\mathbf{k}_{t-B}, \mathbf{k}_{t-B+1},...,\mathbf{k}_{t},\mathbf{k}_{t+1},...,\mathbf{k}_{t+A}]^{\intercal} \frac{\mathbf{q}_{t}}{\sqrt{d_{k}}}.
\end{equation}
Where $\mathbf{z}_{t}$ is now only $A+1+B$ frames in length instead of ${N_{T}}$. $\mathbf{a}_{t}$ is calculated as $softmax(\mathbf{z}_{t})$ as before. For calculating $\mathbf{y}_{t}$, we use

\begin{equation}
	\label{eq:for-v}
	\mathbf{y}_{t} = [\mathbf{v}_{t-B}, \mathbf{v}_{t-B+1},...,\mathbf{v}_{t}, \mathbf{v}_{t+1},...,\mathbf{v}_{t+A}]^{\intercal} \mathbf{a}_{t}.
\end{equation}

We can also express the previous equation in summation form, showing the element-wise multiplications. The Eq.~\eqref{eq:for-v} becomes
\begin{equation}
	\label{eq:element-v}
	y_{i}^{t} = \sum_{j=-B}^{A} v_{i}^{t+j}a_{j+B}^{t}.
\end{equation}

\subsubsection{Back-propagation}
\vspace{-0.1cm}
\label{sec:subsec_derivatives_sa}
As mentioned above, the derivatives of the loss with respect to each of the SDPA inputs need to be computed to update the parameters of the model. This includes $\frac{\partial \ell}{\partial \mathbf{v}_{t}}$, $\frac{\partial \ell}{\partial \mathbf{k}_{t}}$ and $\frac{\partial \ell}{\partial \mathbf{q}_{t}}$, where $\ell$ is the overall loss for one step, corresponding to a mini-batch.

\vspace{-0.25cm}
\subsubsection{Derivative with respect to values}
\vspace{-0.1cm}

By examining Eq.~\eqref{eq:for-v}, we see that
\begin{equation}
	\label{eq:y-v}
	\begin{aligned}
		\frac{\partial \mathbf{y_{n}}}{\partial \mathbf{v_{t}}}=\left \{ \begin{array}{rcl}  
			a_{n, t-n} & \mbox{if } \quad t-A \leq n \leq t+B \\
			0 & \mbox{otherwise},
		\end{array} \right.
	\end{aligned}
\end{equation}
where $\mathbf{a}_{n, t-n}$ denotes the $(t-n)^{th}$ element of $\mathbf{a}_{n}$. $\frac{\partial \ell}{\partial \mathbf{y}_{n}}$ is the input to the back-propagation step.  By applying the chain rule to Eq.~\eqref{eq:y-v}, the full derivative with respect to $\mathbf{v_{t}}$ is:
\vspace{-0.15cm}
\begin{equation}
	\label{eq:L-v-2}
		\frac{\partial \ell}{\partial \mathbf{v}_{t}}=\sum_{n=t-B}^{t+A} a _{n, t-n} \frac{\partial \ell}{\partial \mathbf{y}_{n}} .
\end{equation}

\subsubsection{Derivative with respect to queries}
\noindent The Jacobian matrix of the $softmax(\cdot)$ operator can be found in section 5.3.4 of \cite{bishop2006pattern} and expressed as $\mathrm{J}$.
$\frac{\partial \mathbf{y}_{t}}{\partial \mathbf{a}_{t}}$ can be determined using Eq.~\eqref{eq:dot-product}, and it can be combined with $\mathrm{J}$ to obtain
\begin{equation}
	\label{eq:y-z}
		\frac{\partial \mathbf{y}_{t}}{\partial \mathbf{z}_{t}}=[\mathbf{v}_{t-B}, ...,\mathbf{v}_{t},  \mathbf{v}_{t+1}, ...,\mathbf{v}_{t+A}] \mathrm{J}
	.
\end{equation}
$\frac{\partial \mathbf{z}_{t}}{\partial \mathbf{q}_{t}}$ can be determined using Eq.~\eqref{eq:for-v}. By applying the chain rule to Eq.~\eqref{eq:y-z} and $\frac{\partial \ell}{\partial \mathbf{y}_{n}}$, we obtain
\vspace{-0.15cm}
\begin{equation}
	\label{eq:l-q}
	\frac{\partial \ell}{\partial \mathbf{q}_{t} }=\frac{1}{\sqrt{d_{k}}} \frac{\partial \ell}{\partial \mathbf{y}_{t}} [\mathbf{v}_{t-B}, ...,\mathbf{v}_{t}, ...,\mathbf{v}_{t+A}] \mathrm{J} [\mathbf{k}_{t-B}, ...,\mathbf{k}_{t}, ...,\mathbf{k}_{t+A}]^{\intercal}.
\end{equation}

\subsubsection{Derivative with respect to keys}

From Eq.~\eqref{eq:dot-product}, we obtain
\begin{equation}
	\begin{aligned}
		\label{eq:z-k}
		\frac{\partial \mathbf{z}_{n}}{\partial \mathbf{k}_{t}}=\left \{ \begin{array}{rcl}   \mathrm{M}_{n}  & \mbox{for} \quad t-A \leq n \leq t+B \\ 0 & \mbox{others}, \end{array} \right.
	\end{aligned}
	\vspace{-0.25cm}
\end{equation}
and $\mathrm{M}_{n}$ is a $((A+B+1) \times dk)$ matrix, where the  $(B+t-n)^{th}$ row is specified as $\frac{1}{\sqrt{dk}}\mathbf{q}_{n}^{\intercal}$, and all the other elements are zero. This can be expressed as
\begin{equation}
	\label{eq:m}
	\mathrm{M}_{n}=\left[ \begin{array}{ccc}
		0  &\hdots&0 \\ \vdots & \vdots &\vdots\\q_{n}^{0}&\hdots&q_{n}^{dk}\\ \vdots & \vdots &\vdots\\ 0  & \hdots&0 \end{array} \right]\Bigg\}_{\text{(t-n)}^{th}} .
\end{equation}

Since there is time mixing between $\mathbf{y}$ and $\mathbf{k}$,  we can express the full derivative as
\begin{equation}
	\label{eq:l-k}
	\frac{\partial \ell}{\partial \mathbf{k}_{t}}=\sum_{n=t-A}^{t+B} \frac{\partial \ell}{\partial \mathbf{y}_{n}}\frac{\partial \mathbf{y}_{n}}{\partial \mathbf{z}_{n}} \mathrm{M}_{n}.
\end{equation}
\noindent where $\frac{\partial \mathbf{y}_{n}}{\partial \mathbf{z}_{n}}$ is defined in Eq.~\eqref{eq:y-z}.

\vspace{-0.1cm}
\subsection{Low Latency Streaming Attention}
\vspace{-0.1cm}
\label{sec:lowLatencyStreamingAttentionKernel}
We now introduce Low Latency Streaming Attention (LLSA), a method for preventing the latency buildup due to concatenation of multiple layers of SA, at the expense of higher computational complexity.

\begin{figure}[hbt!]
	\centering
	\begin{minipage}{1.0\textwidth}
		\centering
		\includegraphics[width=0.5\textwidth]{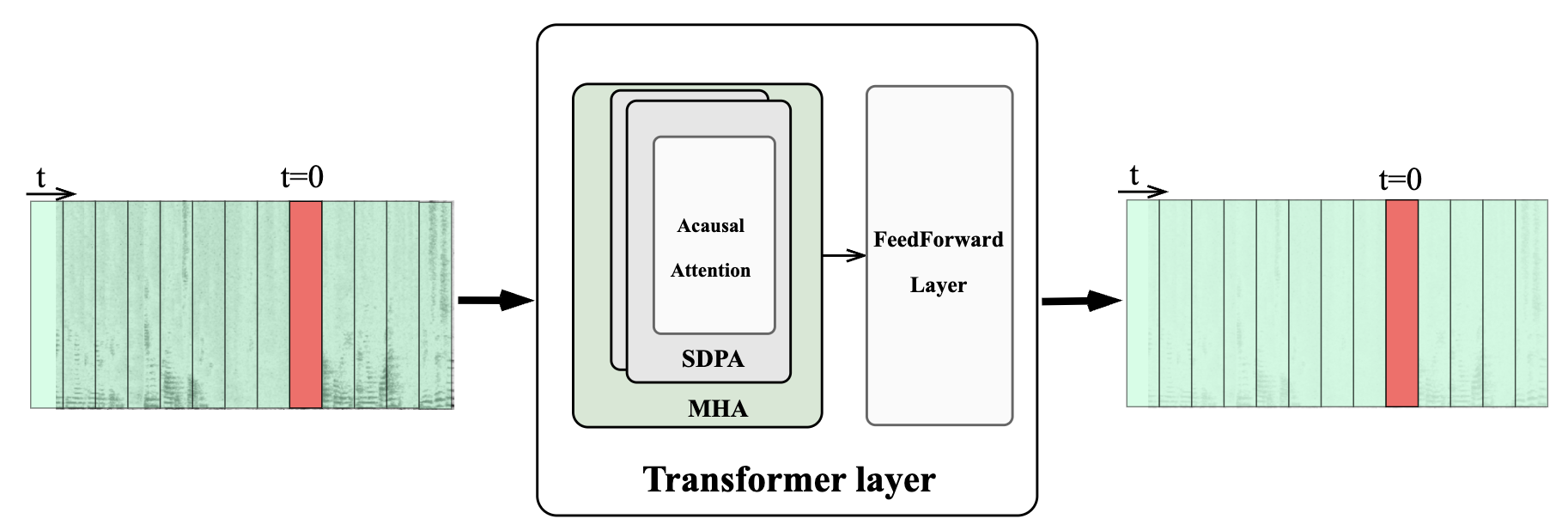}
		\vspace{3pt}
		\label{fig:aa}
	\end{minipage}
	\begin{scriptsize}
		(\textit{a}) Acausal attention (AA), forward function specified in \eqref{eq:na-dot-product} and \eqref{eq:na-for-v}
	\end{scriptsize}
	\begin{minipage}{1.0\textwidth}
		\centering
		\includegraphics[width=0.5\textwidth]{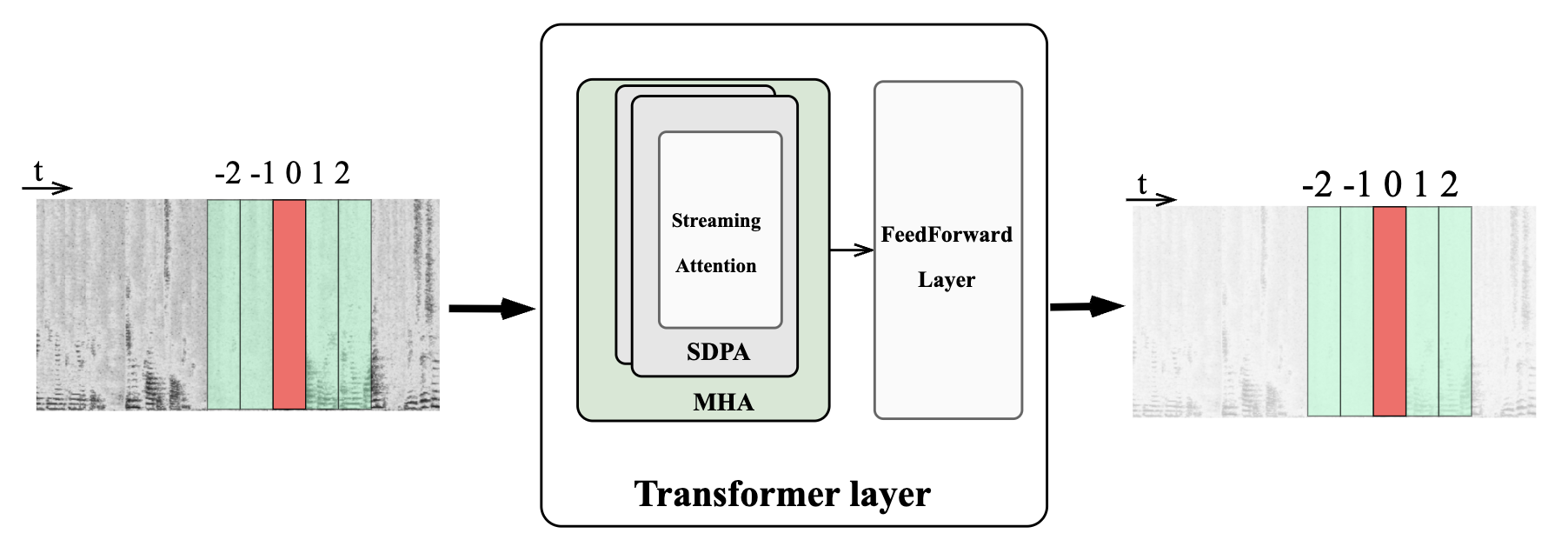}
		\vspace{3pt}
		\label{fig:sa}
	\end{minipage}
	\begin{scriptsize}
		(\textit{b}) Streaming Attention (SA), forward function specified in \eqref{eq:dot-product} and \eqref{eq:for-v}
	\end{scriptsize}
	\begin{minipage}{1.0\textwidth}
		\centering
		\includegraphics[width=0.5\textwidth]{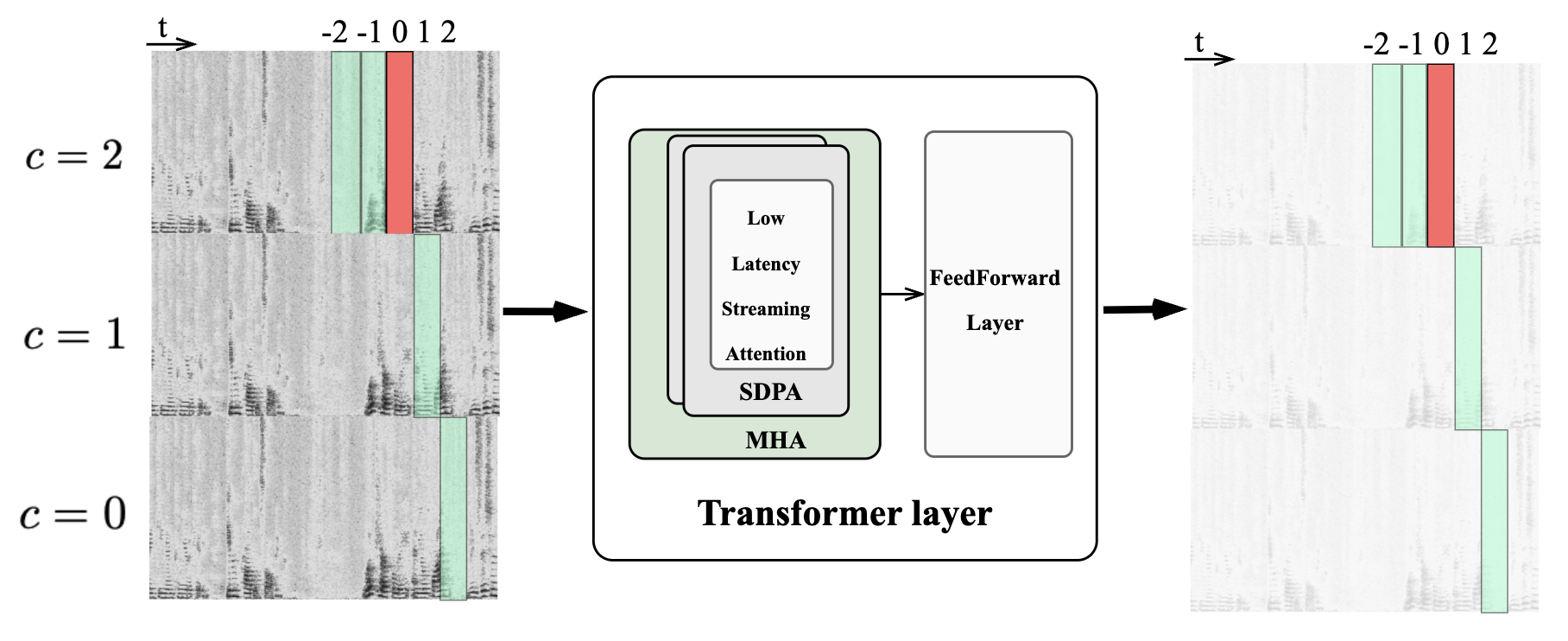}
		\label{fig:llsa}
		\vspace{12pt}
	\end{minipage}
	\begin{scriptsize}
		(\textit{c}) Low Latency SA (LLSA), forward function specified in \eqref{eq:llsa_dot-product} and \eqref{eq:llsa_for-v}
	\end{scriptsize}
	\vspace{0.05cm}
	\caption{Illustration of latency for AA, SA and LLSA. The horizontal axis represents time slice and $t=0$ denotes the current frame. The red box denotes the current query of interest $\mathbf{q}_{t}$ or $\mathbf{q}_{t,c}$ in LLSA. Blue boxes plusing the red box indicate keys and values used in attention. }
	\label{fig:latency_illustration}
\end{figure}

We first define the notation of channel in order to explain the LLSA. In LLSA, a channel is defined as the self-attention uses a specific number of look ahead frames. For example, $c=1$ means inside the SDPA, one look ahead frame is used.  LLSA-based SDPA units then have multiple input channels $\mathbf{q}_{t,c}$, $\mathbf{k}_{t,c}$ and $\mathbf{v}_{t,c}$ as well as multiple output channels $\mathbf{y}_{t,c}$, where each channel of each signal has a unique look-ahead size. To compute the output for each channel, we use a unique $\mathbf{z}_{t,c}$, representing the unnormalised attention score for output channel $c$ at time $t$, and defined as
\vspace{-0.1cm}
\begin{equation}
	\label{eq:llsa_dot-product}
	\begin{aligned}
		\mathbf{z}_{t,c}=[\mathbf{k}_{t-c-B,A},...,\mathbf{k}_{t-c,A}, \mathbf{k}_{t-c+1,A-1}, \cdots,\mathbf{k}_{t+A-c,0}]^{\intercal} \frac{\mathbf{q}_{t,c}}{\sqrt{d_k}}
	\end{aligned}
\end{equation}
\vspace{-0.1cm}
$\mathbf{y}_{t,c}$ is then computed as
\begin{equation}
	\label{eq:llsa_for-v}
	\begin{aligned}
		\mathbf{y}_{t,c}=[\mathbf{v}_{t-c-B,A},...,\mathbf{v}_{t-c,A}, \mathbf{v}_{t-c+1,A-1}, \cdots,\mathbf{v}_{t+A-c,0}]^{\intercal} \mathbf{a}_{t,c}
	\end{aligned}
\end{equation}
where $\mathbf{a}_{t,c}=softmax(\mathbf{z}_{t,c})$.

To compute the derivatives, we employ a procedure analogous to the one outlined in Section \ref{sec:subsec_derivatives_sa}. The detailed calculations will be omitted for brevity,  but the full derivation is available at the end of Section \ref{sec:supplementary}. In the end we obtain
\vspace{-0.1cm}
\begin{equation}
	\label{eq:ll-L-v-2}
	\frac{\partial \ell}{\partial \mathbf{v}_{t,c_2}}=\sum_{c_1} \sum_{n=t-A+c_2}^{t+B+c_2} a_{t-n,n,c_1} \frac{\partial \ell}{\partial \mathbf{y}_{n,c1}},
	\vspace{-0.1cm}
\end{equation}
where $0 \leq {c1, c2} \leq A $ and $a_{t-n,n,c_1}$  denotes the $(t-n)^{th}$ element of $\mathbf{a}_{n,c_1}$. 

Fig. \ref{fig:latency_illustration} gives illustration that compares the input and output of AA, SA and LLSA. As mentioned in Section \ref{sec:aa}, when using AA,  the entire input segment is used to obtain keys and values for each query (Fig. \ref{fig:latency_illustration}(a)). In contrast, Fig. \ref{fig:latency_illustration}(b) shows two layers of transformer with SA. The red box shows the query currently being processed. The green boxes indicate which input frames are used as keys and values when processing that query. In this example, the look-ahead ($A$) and the look-back ($B$) are both set to two frames for each of the two layers. In order to compute the output $\mathbf{y}_{t}$ of the first layer, input at time $t+2$ is required, causing a latency of two frames. For the same reason, the second layer will introduce an additional latency of two frames, giving a total latency of four frames.

Fig. \ref{fig:latency_illustration} (c) shows what can be achieved with LLSA. LLSA prevents latency accumulation by computing multiple output channels at each time step using different look-ahead sizes (except at the beginning of the input where they are simply duplicated). We use $c$ to indicate how many look-ahead frames the version of output uses.  In this example the $c=0$ output is computed using zero look-ahead frames, the $c=1$ output uses one look-ahead frame and the $c=2$ output uses two look-ahead frames. When processing the highlighted red query, all look-back keys and values are extracted from the $c=2$ input, while the look-ahead keys and values are selected from $c=1$ for $t=1$ time index and from $c=0$ for time index $t=2$. When the vector at $c=1, t=1$ is the query, the same keys and values of the red vector are used. This is also true for the vector at $c=0, t=2$.

The motivation of the strategy shown in Fig. \ref{fig:latency_illustration} (c) is more apparent when processing the following layer. When processing the red vector after layer 1, the $c=1,t=1$ and $c=0,t=2$ are already available since they do not depend on additional future vectors other than the five vectors already computed. As a result, the latency does not build up as the number of layers increases.

\subsection{SHuBERT: HuBERT with low latency attention module}
\label{sec:shubertT}
\noindent The proposed low latency attention module is used to replace the vanilla acausal attention module in the transformer block.

The original HuBERT (Fig.~\ref{fig:hubert}) \cite{hsu2021hubert} is characterized by two pre-training iterations. The first iteration uses past and future information to learn a representation of the input data and generate pseudo labels. K-means is used to cluster the output of the first iteration into 500 clusters. The second iteration ingests the pseudo labels from the clustering process and is also acausal. As shown in Fig.~\ref{fig:shubert}, Streaming HuBERT (SHuBERT) shares the same architecture of HuBERT. The main difference is that in SHuBERT we replace the acausal self-attention mechanism of HuBERT with SA, but that happens only during the second pre-training iteration. The main reason for applying SA only at the second pre-training iteration is that it is beneficial to use all the frames available in one utterance to generate pseudo labels in the first pre-training iteration.

After two self-supervised pre-training iterations, ASR is selected as downstream task. This involves replacing the final linear layer with a softmax layer. We use Connectionist Temporal Classification (CTC)~\cite{graves2006connectionist}, adopting the same configuration outlined in \cite{hsu2021hubert}. During the downstream task fine-tuning, we evaluate two distinct configurations. In the first configuration, we fine-tune the model using SA and LLSA, while the second configuration uses LLSA only. The results of our experiments are presented in Section~\ref{sec:experimental_setup}.

\section{Experimental setup}
\label{sec:experimental_setup}
\noindent In this section, the implementation of the low latency attention module is described. Section \ref{sec:datasets} describes the details of the data used to train SHuBERT and the ASR downstream task.  

\subsection{Implementation of low latency attention module}
\label{sec:implementation_lla}
Observing the equations in Section \ref{sec:llam}, we can see that for each frame, there is a different slice of keys, values, unnormalised scores, channels etc. This makes matrix multiplication not applicable and then the use of automated differentiation is inefficient to compute the back-propagation for SA and LLSA architectures. To make the algorithms proposed in this paper efficient, we implemented custom GPU backpropagation CUDA kernels callable by PyTorch \cite{paszke2019pytorch} during off-line model training. Our proposed CUDA SA and LLSA implementations require much less memory than the standard AA implementation. This allows for larger batch sizes during training. Section \ref{sec:experimental_results} includes both theoretical results and simulations showing that the proposed SA implementation is also more memory efficient than the previously proposed masked attention module. While we acknowledge that our SA implementation has not been fully optimized, we are planning to dedicate future work to the creation of optimized SA and LLSA kernels to fully reach the theoretical computational improvement over MAA.

\subsection{Datasets}
\label{sec:datasets}
The HuBERT-base model~\cite{hsu2021hubert} is used as test bed in the experiments. Following the same experimental setup as~\cite{hsu2021hubert}, the train-clean-100, train-clean-360 and train-other of Librispeech~\cite{panayotov2015librispeech} are used for the first and second pre-training iteration of both HuBERT and SHuBERT. We will refer to this training data as librispeech-960. For the ASR fine-tune downstream task, only the train-clean-100 subset is used, following the same process described in~\cite{hsu2021hubert}. Table~\ref{tab:data_usage} shows the data used in the experiments.

\begin{table*}[ht]
	\centering
	\renewcommand\arraystretch{1.5} 
	\begin{tabularx}{\textwidth} { X|X|c } %
		\hline
		\textbf{{ \fontsize{8}{10} \selectfont Stage\quad}} &  \textbf{{ \fontsize{8}{10} \selectfont Subsets}}  & \textbf{{ \fontsize{8}{10} \selectfont Amount (hours)}}  \\
		\hline
		\fontsize{7}{10} \selectfont First iteration  & \fontsize{7}{10} \selectfont librispeech-960 & \fontsize{7}{10} \selectfont 960 \\ 
		\hline
		\fontsize{7}{10} \selectfont Second iteration  & \fontsize{7}{10} \selectfont librispeech-960 & \fontsize{7}{10} \selectfont 960 \\ 
		\hline
		\fontsize{7}{10} \selectfont ASR fine-tune  & \fontsize{7}{10} \selectfont train-clean-100 & \fontsize{7}{10} \selectfont 100 \\ 
		\hline
		\fontsize{7}{10} \selectfont Development  & \fontsize{7}{10} \selectfont dev-clean, dev-other & \fontsize{7}{10} \selectfont 10.7 \\ 
		\hline
		\fontsize{7}{10} \selectfont Test  & \fontsize{7}{10} \selectfont test-clean, test-other, & \fontsize{7}{10} \selectfont 10.5 \\ 
		\hline
	\end{tabularx}
	\caption{Datasets used in the experiments.}
	\label{tab:data_usage}
\end{table*}

\section{Experimental Results}
\label{sec:experimental_results}
We ran two distinct experiments. We first evaluated the computational efficiency of our implementation, and subsequently we quantified the performance of Streaming HuBERT (SHuBERT) in a streaming self-supervised speech representation learning use case. 

\subsection{Computational Efficiency}
\label{sec:computational_efficiency}
As mentioned in Section \ref{sec:implementation_lla}, in this section we present the theoretical computational performance of our implementation, followed by the empirical experimental results.
\subsubsection{Lower Compute Bound}
In Section \ref{sec:subsec_mask}, we reported the theoretical complexity of Masked Acausal Attention (MAA) as $O(d_{k} N_{T}^{2})$, which was estimated using redundant dummy value replacements. In comparison, the computational complexity of our approach (SA) is equal to $O(d_{k} N_{T}*(A+B+1))$, where A refers to the number of look-ahead vectors and B is the number of look-back vectors. This shows that the theoretical complexity of SA is significantly reduced. 

While our existing implementation may not achieve the anticipated computational improvement, we do observe a linear correlation between compute trends and receptive field size, affirming our theoretical approximation. The failure to attain the theoretical computational gain is likely attributed to a lack of code optimization. We opt not to present experimental results on execution time, since we believe this would be an unfair comparison against the highly optimized CUDA operators in PyTorch, like the MMA implementation.

\subsubsection{Lower Memory Usage}
In MAA, many matrices are stored in memory for forward and backward propagations. This includes mask matrices and intermediate vectors like $\mathbf{z_t}$, $\mathbf{a_t}$. Those vectors are $N_T \times N_T$ in size and they are stored for long time periods, causing a large impact on total memory requirement. By contrast, the attention scores in SA (such as $\mathbf{z_t }$ or $\mathbf{a_t }$) have a size of $N_T \times (A+B+1)$. In applications where the length of the vectors is much longer than the receptive field of the SA transformer ($A+B+1$), a significant theoretical memory saving is expected. For example, if $N_T$ is in the order of 3000 (30s sequence with 10ms hop), we expect a receptive field for SA of $\sim$300 .

\begin{figure}
	\centering
	\includegraphics[scale=0.5]{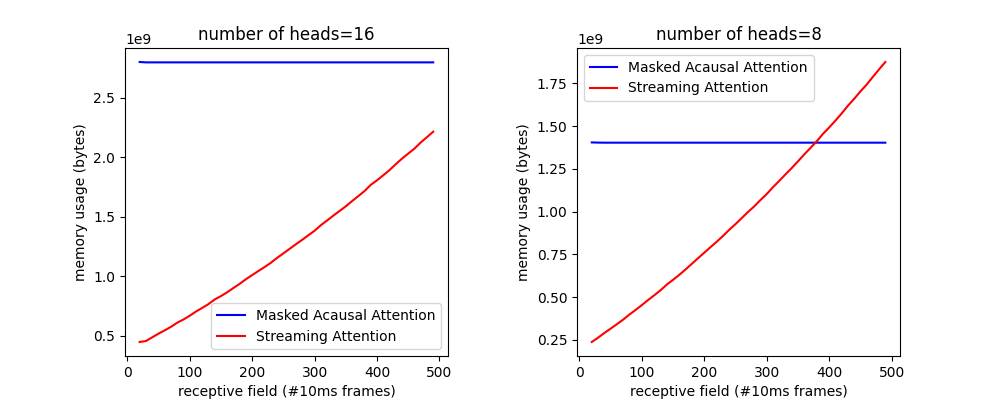}
	\caption{Validation of memory efficiency. \textit{Test platform}: GPU GeForce RTX 2080, Intel® Xeon®(R) Gold 6144 CPU @ 3.50GHz, PyTorch (1.9.0+cu111)}
	\label{fig:memory_efficiency}
	\vspace{10pt}
\end{figure}

In addition to this theoretical computation, Fig. \ref{fig:memory_efficiency} shows the memory profiling of the proposed SA, highlighting this advantage.
In this test experiment, we used a configuration of the attention that is frequently used in many applications. The specifications are as follows: the dimension of each attention is $d_k=64$, the number of frames for the input is $N_T=1000$, the number of heads in the attention mechanism varies from $(8, 16)$, and the size of the receptive field $(A+B+1)$ varies from 10 to 490 with a step of 10. The experiments were repeated 5 times, and the mean values are reported.

In Fig. \ref{fig:memory_efficiency}, the red line denotes the memory usage for SA and the blue line is the memory usage for MAA. The memory usage for MAA is constant because it does not depend on the size of the receptive field. The results show that our method requires less memory per training vector than the MAA approach in many useful configurations, particularly when the number of heads is 8 or 16. The advantage is more pronounced when the receptive field is shorter, as expected. The results show that SA requires less memory per training vector both practically and theoretically, enabling to use a larger batch size for a given GPU memory size. As additional anecdotal validation of this experimental result, we were able to train our model using a single GPU, while 2 GPUs are required to train MAA with an equivalent batch size.

\subsection{Experiments of SHuBERT}
\label{sec:SHuBERT_result}
The experiments are broadly divided into two categories. The first one is the offline model which is used as baseline to compare. The second one is the streaming version (SHuBERT). Both used the same datasets as indicated in Section~\ref{sec:datasets}. Experimental setups such as optimizers, learning rate, and batch-size are the same across them. The number of training steps is the same for all experiments except the one using SA and LLSA in the fine-tune part to verify the effectiveness of LLSA (which indicated as UPsa\_FTsallsa in Table~\ref{tab:shubert_asr_task}). Other than that, the only difference of experimental setup for these two categories is that in SHuBERT, the SA and/or LLSA are used in the second pre-training iteration and fine-tune part, and the conventional AA is used in HuBERT.

To explore the effectiveness of the SA and LLSA method in fine-tune part, we further conduct experiments that compare three different cases. They are: using SA only; using LLSA only; using SA first and then LLSA with varying step number. As LLSA typically is more computationally expensive, SA is only used in the pre-training part as it can train with more data within the same time.

Other experimental setups follow the same as the HuBERT-base in~\cite{hsu2021hubert}. Two iterations of pre-training with 960 hours of LibriSpeech audio with minicing 32GPUs, with overall batch size of at most 2800 seconds of audio. The number of steps for first iteration is 250k steps and it 400k for the second iteration. Pseudo labels for the second iteration are generated by clustering the 6-th transformer layer output of the first iteration model. The mask probability is also the same as~\cite{hsu2021hubert}, which is set to $l= 10$, and $p=8\%$. Adam~\cite{kingma2014adam} optimizer is used with $\beta=(0.9,0.98)$, and the same learning rate schedue for HuBERT-base~\cite{hsu2021hubert} is used. The peak learning rate is 5e-4.

It is worth noting that for SHuBERT, we can use SA or LLSA during decoding. This is denoted as infer\_sa using SA and infer\_llsa using LLSA during decoding in Table~\ref{tab:shubert_asr_task}. The method infer\_sa requires larger latency and the latency is denoted in fourth column in Table~\ref{tab:shubert_asr_task}, where it can be seen the theorectial latency for infer\_llsa are significantly reduced.

\subsubsection{Experimental results}
As mentioned before, the first category is the baseline HuBERT-base model. It was trained by following the same procedure described in \cite{hsu2021hubert}. We subsequently fine-tune both models for the downstream ASR task on librispeech-clean-100h, as others have done in the literature~\cite{huang2023self}\cite{yang2021superb}\cite{chang2022distilhubert}. The results are shown in the second row of table~\ref{tab:hubert_asr_baselines}. It can be seen that our implementation performs comparably to the numbers reported in \cite{hsu2021hubert}, especially on test-clean.
\begin{table*}[!ht]
	\centering
	\begin{tabularx}{\columnwidth} { l|c|c|c|c } %
		\hline
		\textbf{{ model\quad}} & \textbf{{ dev-clean}} & \textbf{{ dev-other}} & \textbf{{ test-clean}} & \textbf{{ test-other}}  \\
		\hline
		\fontsize{7}{10} \selectfont HuBERT-base \cite{hsu2021hubert} & 2.70 & 7.80 & 3.40 & 8.10 \\  
		\hline
		\fontsize{7}{10} \selectfont HuBERT (ours) & 2.80 & 8.32 & 3.47 & 8.24 \\  
		\hline
	\end{tabularx}
	\caption{ASR Baselines Results (Word Error Rate (WER) \% ) }
	\label{tab:hubert_asr_baselines}
\end{table*}

As mentioned before, we conducted the ASR downstream task on SHuBERT model and the results are shown in table~\ref{tab:shubert_asr_task}. Similar to other recent work to convert an SSL model to its streaming version, the SA is applied during the second iteration of the upstream model training, which we denote as UPsa in table~\ref{tab:shubert_asr_task}. In order to thoroughly analyze the configurations, three sets of experiments were conducted by varying which attention module was used.
\begin{table}[!ht]
		\begin{tabular}{l ccccccc}
			\toprule
			\textbf{{ model}} & \textbf{{ conf.}} & \textbf{{infer type}} & \textbf{{latency}} & \textbf{{dev-clean}} & \textbf{{dev-other}} & \textbf{{test-clean}} & \textbf{{test-other}}  \\
			\midrule
			\multirow{4}{*}{{ \fontsize{ 7 }{10} \selectfont UPsa\_FTsa }} &  \fontsize{ 6 }{10} \selectfont l32\_r8 & \multirow{2}{*}{\fontsize{ 6 }{10} \selectfont infer\_sa} & \fontsize{ 6 }{10} \selectfont 1.92s & \fontsize{ 6 }{10} \selectfont 4.01 & \fontsize{ 6 }{10} \selectfont 14.13 & \fontsize{ 6 }{10} \selectfont 4.51 & \fontsize{ 6 }{10} \selectfont 14.84 \\
			\cmidrule(lr){2-2} \cmidrule(lr){4-8}
			& \fontsize{ 6 }{10} \selectfont l32\_r16 & & \fontsize{ 6 }{10} \selectfont 3.84s & \fontsize{ 6 }{10} \selectfont 3.97 & \fontsize{ 6 }{10} \selectfont 13.80	 & \fontsize{ 6 }{10} \selectfont 4.26 & \fontsize{ 6 }{10} \selectfont 14.33 \\
			\cmidrule(lr){2-8}
			&  \fontsize{ 6 }{10} \selectfont l32\_r8 & \multirow{2}{*}{\fontsize{ 6 }{10} \selectfont infer\_llsa} & \fontsize{ 6 }{10} \selectfont 0.16s & \fontsize{ 6 }{10} \selectfont 12.99 & \fontsize{ 6 }{10} \selectfont \fontsize{ 6 }{10} \selectfont 36.18 & \fontsize{ 6 }{10} \selectfont 13.82 & \fontsize{ 6 }{10} \selectfont 31.82 \\
			\cmidrule(lr){2-2} \cmidrule(lr){4-8}
			& \fontsize{ 6 }{10} \selectfont l32\_r16 & & \fontsize{ 6 }{10} \selectfont 0.32 & \fontsize{ 6 }{10} \selectfont 6.83 & \fontsize{ 6 }{10} \selectfont \fontsize{ 6 }{10} \selectfont 23.72 & \fontsize{ 6 }{10} \selectfont 7.41 & \fontsize{ 6 }{10} \selectfont 25.07 \\
			\midrule
			\multirow{2}{*}{{ \fontsize{ 7 }{10} \selectfont UPsa\_FTsallsa }} &  \fontsize{ 6 }{10} \selectfont l32\_r8 & \multirow{2}{*}{\fontsize{ 6 }{10} \selectfont infer\_llsa} & \fontsize{ 6 }{10} \selectfont 0.16s & \fontsize{ 6 }{10} \selectfont 5.16 & \fontsize{ 6 }{10} \selectfont \fontsize{ 6 }{10} \selectfont 18.73 & \fontsize{ 6 }{10} \selectfont 5.82 & \fontsize{ 6 }{10} \selectfont 19.45 \\
			\cmidrule(lr){2-2} \cmidrule(lr){4-8}
			& \fontsize{ 6 }{10} \selectfont l32\_r16 & & \fontsize{ 6 }{10} \selectfont 0.32s & \fontsize{ 6 }{10} \selectfont 4.62 & \fontsize{ 6 }{10} \selectfont \fontsize{ 6 }{10} \selectfont 15.98 & \fontsize{ 6 }{10} \selectfont 5.02 & \fontsize{ 6 }{10} \selectfont 17.00 \\
			\midrule
			\multirow{2}{*}{{ \fontsize{ 7 }{10} \selectfont UPsa\_FTllsa }} &  \fontsize{ 6 }{10} \selectfont l32\_r8 & \multirow{2}{*}{\fontsize{ 6 }{10} \selectfont infer\_llsa} & \fontsize{ 6 }{10} \selectfont 0.16s & \fontsize{ 6 }{10} \selectfont 5.31 & \fontsize{ 6 }{10} \selectfont \fontsize{ 6 }{10} \selectfont 18.76 & \fontsize{ 6 }{10} \selectfont 5.84 & \fontsize{ 6 }{10} \selectfont 19.46 \\
			\cmidrule(lr){2-2} \cmidrule(lr){4-8}
			& \fontsize{ 6 }{10} \selectfont l32\_r16 & & \fontsize{ 6 }{10} \selectfont 0.32s & \fontsize{ 6 }{10} \selectfont 4.68 & \fontsize{ 6 }{10} \selectfont \fontsize{ 6 }{10} \selectfont 15.49 & \fontsize{ 6 }{10} \selectfont 4.98 & \fontsize{ 6 }{10} \selectfont 16.55 \\
			\bottomrule
		\end{tabular}
		\caption{SHuBERT ASR Downstream Task Results (WER \% ). The UPsa denotes the upstream HuBERT model used the SA and FT denotes downstream task fine-tune process. The annotation after FT denotes which attention module is used e.g. FTsallsa denotes SA is first used and then LLSA is applied after SA, while FTllsa means only llsa is used. }
		\label{tab:shubert_asr_task}
\end{table}
In the first set of experiments, where the UPsa\_FTsa model was analyzed, the SA is used during downstream ASR task fine tuned with $30$ thousand (K) updating steps until the CTC loss flattened.  As the SA module has the same functionality as MAA but with better memory usage and less computational burden, it serves as the baseline of SHuBERT-streaming ASR task. Two different latency configurations are implemented. l32\_r8 implies that within the streaming version of self-attention, 32 history frames and 8 look ahead frames were used, and infer\_sa denotes the SA module is used during inference. Due to the latency build-up for infer\_sa, the latency is larger than infer\_llsa. Although SA is only used in the training procedure of UPsa\_FTsa, the LLSA can be still used during decoding. However, the performances are expected to drop as there is mismatch between training and decoding. From Table~\ref{tab:shubert_asr_task}, it can be seen without fine-tune with LLSA, the performance degrades sharply when reducing the latency using the LLSA inference. This trend is consistent for both latency configurations, though larger latency is affected less. It also aligns with the expectation that larger latency generally performs better in terms of WER.

Comparing with UPsa\_FTsa, UPsa\_FTllsa directly uses LLSA for downstream task fine-tune with the same $30$K updating steps. The performance gaps when reducing latency during inference with LLSA are largely closed by the LLSA training. This shows the effectiveness of the LLSA training in order to reduce latency of applications in real-time inference. However, as indicated in Section~\ref{sec:lowLatencyStreamingAttentionKernel}, the LLSA needs to conduct extra $A$ times more computations. This is denoted by the two extra frame vectors in Fig.~\ref{fig:latency_illustration} (C). This observation motivates the third set of experiments denoted as UPsa\_FTsallsa where SA is first applied during the downstream task fine-tune with $25$K updates, and extra $15$K updates with LLSA. The results shows that with help of SA fine-tune which is more computationally efficient, comparable performance can be achieved by only using half of the updating steps with LLSA.

\begin{figure}
	\centering
	\includegraphics[scale=0.5]{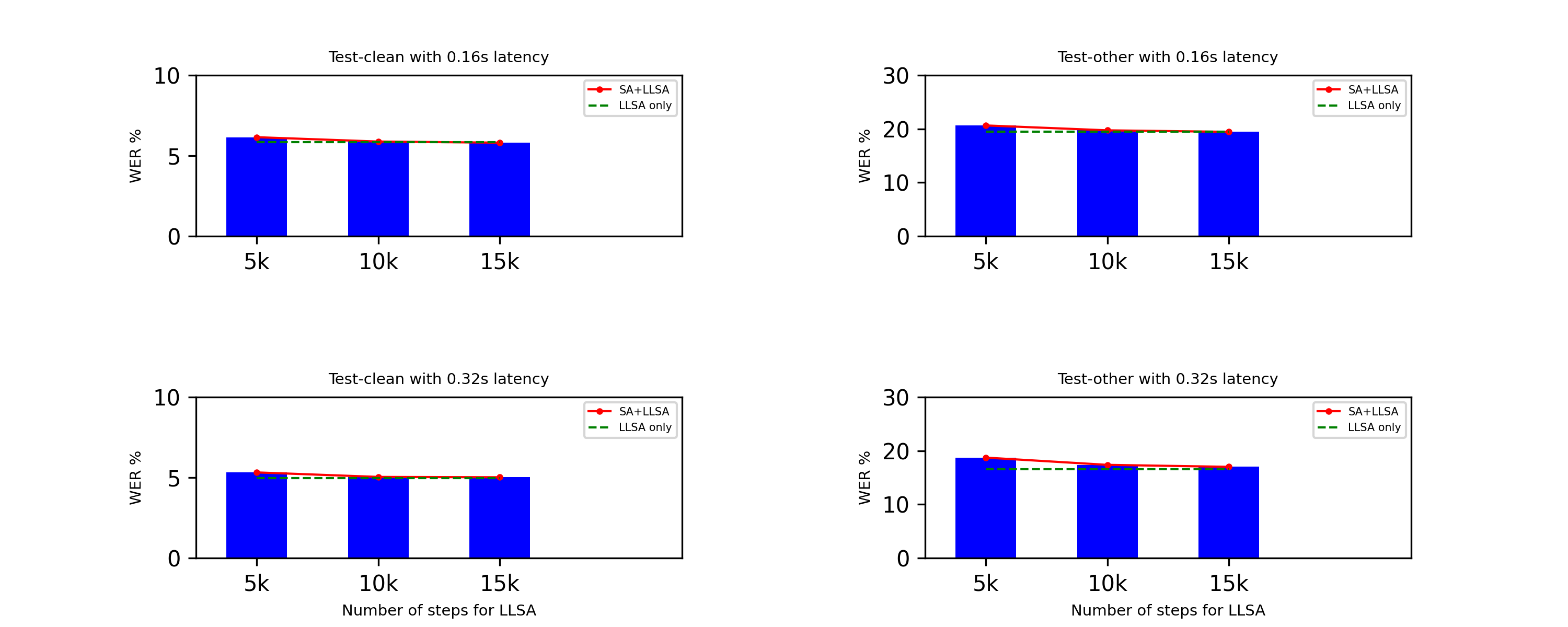}
	\caption{Ablation study of the updating steps of LLSA.}
	\label{fig:SA_LLSA}
	\vspace{-10pt}
\end{figure}

It is possible to make a trade-off between the performance and training efficiency by varying the number of updating steps of LLSA during fine-tune. Fig.~\ref{fig:SA_LLSA} gives the detailed study by varying the number of updating steps. It can be seen that generally larger training steps of LLSA in UPsa\_FTsallsa can achieve more matching performance compared with UPsa\_FTllsa. However, it is also indicated that 15K steps (50\% of the updating steps UPsa\_FTllsa) has already reached comparable performance.

\section{Conclusion}\label{sec:conclusion}

We introduce a new class of low-latency attention module which can be used in transformers, that overcomes the main limitation of the traditional attention mechanism: the acausality. Our solution builds on past work on causal self-attention masking, improving upon computational complexity, memory usage and latency. To achieve this, we propose {\it Streaming Attention (SA)}, a method which increases efficiency and reduces computation redundancy of causal self-attention masks, and {\it Low-Latency Streaming Attention (LLSA)}, which prevents latency accumulation across transformer layers.

In this paper, we show a reduction of theoretical complexity and memory usage over traditional self-attention masking. The proposed low-latency attention module is applied to the HuBERT model to obtain a streaming version of HuBERT (SHuBERT), which performs competitively on the ASR downstream task with only 100 hours of labelled data. SA reduces memory usage during self-supervised training, and LLSA enables a reduction of latency by more than 10 folds (from 3.84 seconds to 0.32 seconds) with only 16.90\% WER drop.

While we have shown applicability of our technology to ASR, we believe its applicability can be extended to support additional downstream tasks, including real-time noise suppression and talker identification, and is not limited to the model architectures covered in this paper, but can be extended to most transformer-based models. 

In conclusion, our Streaming Attention (SA) and Low Latency Streaming Attention (LLSA) techniques provide efficient ways to enable causality in transformer architectures, which is important for processing streaming audio with fixed latency. Our solution opens up the possibility to use transformer-based architectures in new scenarios such as telecommunication, broadcasting, and other real-time applications.  

\backmatter

\bmhead{Supplementary information}
\label{sec:supplementary}



The detailed development of the low latency attention module is provided in the supplementary and can be found in the \href{https://github.com/jianboma/Low_latency_attention_module/blob/main/doc/low_latency_attention_module.pdf}{url}.


\bibliography{literatures}

\end{document}